\documentclass[%
 reprint,
superscriptaddress,
 amsmath,amssymb,
 aps,
pra,
floatfix,
]{revtex4-2}

\usepackage{graphicx}
\usepackage{dcolumn}
\usepackage{bm}

\usepackage{hyperref}
\usepackage[utf8]{inputenc}
\usepackage[english]{babel}

\usepackage{amsfonts}
\usepackage{algorithmic}
\usepackage{textcomp}
\usepackage{xcolor}
\usepackage{multirow}
\usepackage{braket}
\usepackage{quantikz}
\usepackage{tikz}
\usepackage{graphicx}
\usepackage{totpages}
\usepackage{cleveref}
\usepackage{siunitx}
\renewcommand{\selectlanguage}[1]{} 

\Crefname{equation}{Eq.}{Eqs.}
\crefname{equation}{Eq.}{Eqs.}
\Crefname{definition}{Def.}{Defs.}
\crefname{definition}{Def.}{Defs.}
\Crefname{figure}{Fig.}{Figs.}
\crefname{figure}{Fig.}{Figs.}
\Crefname{section}{Sec.}{Secs.}
\crefname{section}{Sec.}{Secs.}
\Crefname{assumption}{Assum.}{Assums.}
\crefname{assumption}{Assum.}{Assums.}
\Crefname{appendix}{Appendix}{Appendices}
\crefname{appendix}{Appendix}{Appendices}

\AddToHook{cmd/appendix/before}{%
  \crefalias{section}{appendix}%
  \crefalias{subsection}{subappendix}%
  \crefalias{subsubsection}{subsubappendix}%
}

\begin{document}

\preprint{APS/123-QED}

\title{Efficient Routing of Quantum LDPC Codes on Programmable 2D Toric Architectures}

\author{Kun Liu}
\email{kun.liu.kl944@yale.edu}
\affiliation{Department of Computer Science, Yale University, New Haven, Connecticut 06511, USA}
\affiliation{Yale Quantum Institute, Yale University, New Haven, Connecticut 06511, USA}

\author{Takahiro Tsunoda}
\affiliation{Department of Applied Physics, Yale University, New Haven, Connecticut 06511, USA}
\affiliation{Yale Quantum Institute, Yale University, New Haven, Connecticut 06511, USA}

\author{Sophia H. Xue}
\affiliation{Department of Applied Physics, Yale University, New Haven, Connecticut 06511, USA}
\affiliation{Yale Quantum Institute, Yale University, New Haven, Connecticut 06511, USA}

\author{Evan McKinney}
\affiliation{Department of Computer Science, Yale University, New Haven, Connecticut 06511, USA}
\affiliation{Yale Quantum Institute, Yale University, New Haven, Connecticut 06511, USA}
\affiliation{Department of Applied Physics, Yale University, New Haven, Connecticut 06511, USA}

\author{Zeyuan Zhou}
\affiliation{Department of Computer Science, Yale University, New Haven, Connecticut 06511, USA}
\affiliation{Yale Quantum Institute, Yale University, New Haven, Connecticut 06511, USA}

\author{Shifan Xu}
\affiliation{Department of Applied Physics, Yale University, New Haven, Connecticut 06511, USA}
\affiliation{Yale Quantum Institute, Yale University, New Haven, Connecticut 06511, USA}

\author{Robert J. Schoelkopf}
\affiliation{Department of Applied Physics, Yale University, New Haven, Connecticut 06511, USA}
\affiliation{Yale Quantum Institute, Yale University, New Haven, Connecticut 06511, USA}
\affiliation{Department of Physics, Yale University, New Haven, Connecticut 06511, USA}

\author{Yongshan Ding}
\email{yongshan.ding@yale.edu}
\affiliation{Department of Computer Science, Yale University, New Haven, Connecticut 06511, USA}
\affiliation{Yale Quantum Institute, Yale University, New Haven, Connecticut 06511, USA}
\affiliation{Department of Applied Physics, Yale University, New Haven, Connecticut 06511, USA}

\date{\today}

\begin{abstract}

Quantum low-density parity-check codes are promising candidates towards scalable fault-tolerant quantum computation. Among these, bivariate bicycle (BB) codes offer superior encoding rates and large code distance compared to surface codes. However, their requirement on long-range stabilizer measurements poses significant challenges for implementation on realistic hardware with limited connectivity, such as superconducting circuit platforms. In this work, we introduce a novel hardware-software co-design that leverages a programmable communication network architecture to address these limitations. Our approach utilizes a 2D toric network of oscillators as a flexible communication fabric linking qubits at each site. Such architecture significantly reduces the number of long-range couplers required from $O(n)$ to $O(\sqrt{n})$. Dual-rail qubits, along with native gates including Swap-Wait-Swap gates and beamsplitter SWAPs, ensure that long-range two-qubit gates can be executed with high fidelity and low latency. To further enhance performance, our qubit layout and routing algorithm utilize symmetries of the codes and enable maximum parallelism for long-range two-qubit gates, maintaining a low syndrome extraction cycle duration and scalability over the code length. We perform circuit-level simulation with realistic noise modeling based on experimental hardware parameters, observing an logical error rate per logical qubit per cycle of 3.06\% for $[[18, 4, 4]]$ BB code, 2.6$\times$ less than the existing experimental result. These findings provide a practical roadmap and identify key technological advancements needed to achieve low-overhead fault-tolerant quantum computing at scale.

\end{abstract}
\maketitle


\section{Introduction}

To execute large-scale quantum algorithms that are classically intractable, hardware-efficient quantum error correction is indispensable.
Today's quantum hardware is imperfect (e.g., physical error rates around \(10^{-3}\) per gate~\cite{beverland_assessing_2022}) and still limited to only hundreds of qubits.
Quantum error correction (QEC) codes protect quantum information by redundantly encoding it across many physical qubits to create logical qubits that are robust against errors.

The surface codes~\cite{bravyi_quantum_1998, dennis_topological_2001, kitaev_fault-tolerant_2003} are widely regarded for their architectural simplicity, compatibility with weight-4 check operators, and a high error threshold of about 1\%~\cite{fowler_high_2012}.
These properties have enabled numerous experimental realizations on superconducting qubit platforms~\cite{krinner_realizing_2022, marques_logical-qubit_2021, sivak_real-time_2023, takita_demonstration_2017, zhao_realization_2022}.

However, the major drawback of surface codes is their low encoding rate, which demands a vast number of physical qubits to host only a few logical qubits.
This inefficiency severely limits scalability under current fabrication constraints; for example, a 5k-qubit superconducting chip, even at an optimistic 99.98\% junction yield, would on average succeed just once in three thousand fabrication attempts~\cite{yoder_tour_2025}.
It is therefore natural to search for QEC codes with better encoding efficiency while remaining compatible with realistic hardware.

The Bravyi-Poulin-Terhal bound~\cite{bravyi_tradeoffs_2010} shows that, under purely local interactions on a 2D lattice, surface codes are optimal in encoding rate.
Relaxing the 2D locality requirement unlocks dramatic improvements through quantum low-density parity-check (qLDPC) codes~\cite{gottesman_fault-tolerant_2014,panteleev_degenerate_2021, panteleev_quantum_2022, tillich_quantum_2014}, which maintain constant-weight stabilizers yet permit long-range interactions, leading to asymptotically constant encoding rates.

Bivariate bicycle (BB) codes~\cite{bravyi_high-threshold_2024} are a finite-length qLDPC family that outperforms surface codes, featuring weight-6 stabilizer checks, a theoretical pseudo-threshold near 0.7\%, and over 10$\times$ savings in physical qubits.
Beyond serving as logical memories, qLDPC codes now enhance logical operation schemes~\cite{cowtan_fast_2025, cowtan_parallel_2025, cross_improved_2024, he_extractors_2025, williamson_low-overhead_2024}, and IBM's bicycle architecture~\cite{yoder_tour_2025} demonstrates that BB codes can deliver 16--17$\times$ more logical qubits and up to 5.8$\times$ more T-gate capacity
(number of maximally affordable non-Clifford rotations) for the same physical-qubit budget and physical error rate compared to surface code systems.
For application circuits such as transverse-field Ising model simulations, the bicycle architecture has been projected to require 2--9$\times$ fewer physical qubits at comparable noise levels~\cite{yoder_tour_2025}.

These projections, however, assume uniform physical error rates across all operations, which is challenging to realize because BB codes' stabilizers are inherently non-local.
BB codes require long-range two-qubit (2Q) gates between qubits typically separated by 10--30 lattice spacings, driving up hardware complexity, crosstalk, and latency.
In the fixed-coupler design~\cite{bravyi_high-threshold_2024}, the number of long-range couplers scales as \(O(n)\) while the interaction range scales as \(O(\sqrt{n})\), creating prohibitive implementation overheads.

Existing superconducting platforms are constrained to nearest-neighbor connectivity, so emulating long-range gates through chains of SWAPs or Bell-pair teleportation~\cite{baumer_efficient_2023, berthusen_toward_2024} compounds noise.
At the time of writing, the only BB code experiment implements the $[[18,4,4]]$ code with 32 long-range couplers and an average CZ error rate of 0.98\%, whereas the more competitive $[[144,12,12]]$ code would demand 324 long-range couplers with maximum range around 23 lattice spacings, making near-term demonstrations highly challenging.

We address these challenges with a hardware-software co-design tailored to superconducting toric architectures.
Our approach introduces a \textbf{programmable communication network} that uses a 2D torus of oscillators as a reconfigurable communication fabric connecting data and check qubits located at each site.
This fabric leverages dual-rail qubits, Swap-Wait-Swap (SWS) entangling gates, and beamsplitter SWAPs to shuttle photons so that a long-range CZ can be executed with high fidelity and low latency.

By reducing the number of fixed long-range couplers from \(O(n)\) to \(O(\sqrt{n})\), the architecture dramatically simplifies hardware requirements while preserving support for complex long-range interaction patterns.
We complement the hardware with a symmetry-aware qubit layout and routing algorithm that exploit the translation- and cell-invariant structure of BB codes.
These optimizations enable \textbf{maximum parallelism}: all 2Q gates required within each layer of the syndrome-extraction circuit execute simultaneously, maintaining short cycle times as the code scales.
Because connectivity can be reconfigured dynamically instead of being fixed during fabrication, the same platform can adapt to different qLDPC codes and potentially support logical operations.

We perform realistic noise modeling based on experimental dual-rail qubit
and SWS gate parameters and run circuit-level simulations, observing a logical error rate per logical qubit per cycle of 3.06\% for the $[[18,4,4]]$ BB code, a 2.6$\times$ improvement over the best published experiment.
These results chart a concrete roadmap for implementing high-rate qLDPC codes on superconducting hardware and illuminate the key technological advances needed for low-overhead fault-tolerant quantum computing.

The remainder of this paper is organized as follows.
\Cref{sec:background} reviews BB codes and prior implementations.
\Cref{sec:approach} introduces the programmable communication network architecture.
\Cref{sec:layout_and_routing} details the qubit layout and routing strategy that yield maximum parallelism.
\Cref{sec:error_modeling} presents the hardware-informed error model.
\Cref{sec:evaluation} reports our simulation and evaluation results.

\begin{figure}[t]
    \centering
    \includegraphics[width=\linewidth, trim={0 1em 0 0}, clip]{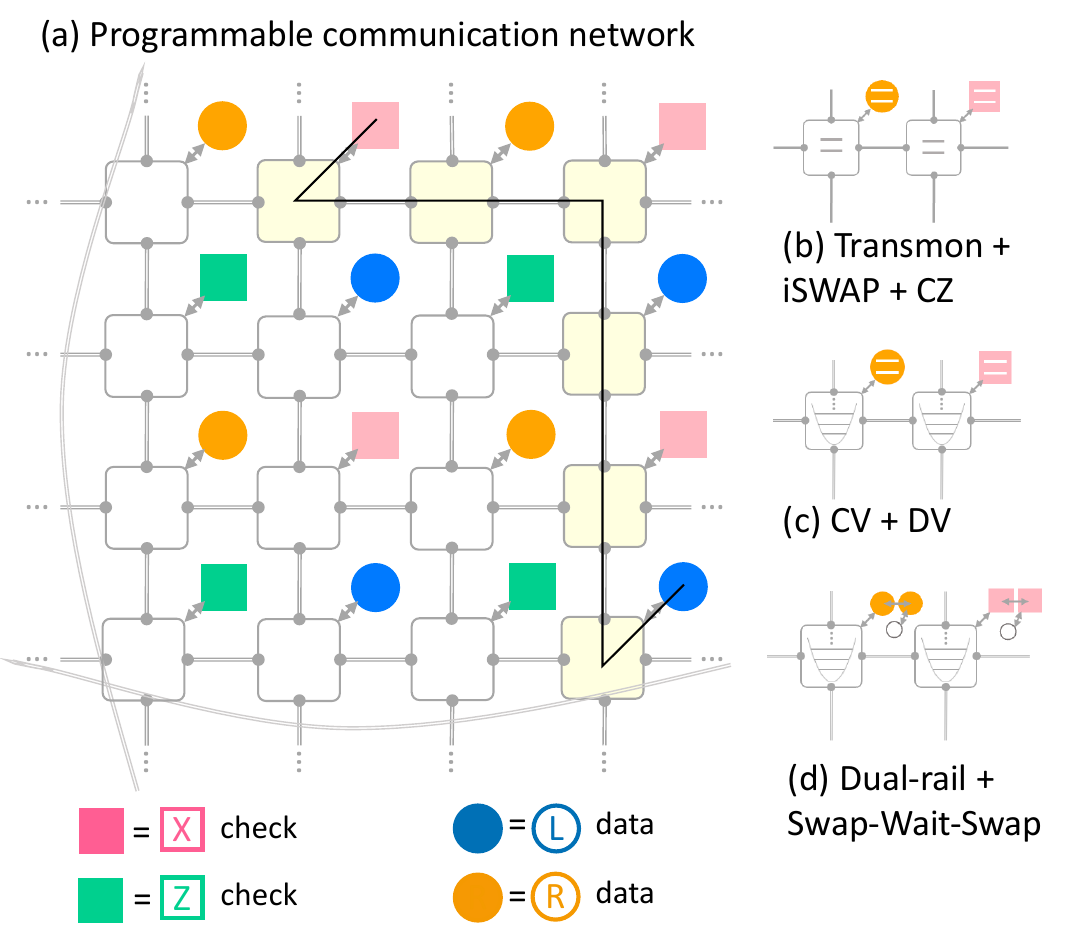}
    \caption{(a) Programmable communication network.
    Communication mediums are connected via physical couplers, allowing quantum states to be efficiently transferred between distant qubits without the need for direct qubit-to-qubit coupling.
    (b--d) are different instances of the programmable communication network.
    (b) Both the medium and the qubit are made of transmons.
    (c) The medium is made of oscillators or cavities, (Continuous Variable, CV) and the qubit is made of transmons (Discrete Variable, DV).
    (d) The medium is made of oscillators (cavities) and the qubit is made of dual-rail qubits. More details in \Cref{sec:approach}.
    }
    \label{fig:programmable_comm_network}
\end{figure}

\section{Background and Motivation}
\label{sec:background}

\subsection{Theory of bivariate bicycle (BB) codes}
\label{sec:bb-codes}

A BB code is defined by two polynomials
$A = A_1 + A_2 + A_3$ and $B = B_1 + B_2 + B_3$,
where each \( A_i \) and \( B_i \) is described by a distinct monomial of the form $x^p y^q$, $ 0 \leq p < \ell$ and $ 0 \leq q < m $.
The parity-check matrices are
$H^X = [A|B]$ and $H^Z = [B^T|A^T]$,
where \( H^X \) and \( H^Z \) are $n/2 \times n$ matrices with \( n = 2 \ell m \).
We refer to the first $n/2$ data qubits as $L$ qubits, the second $n/2$ as $R$ qubits, the rows of \( H^X \) as $X$ checks, and the rows of \( H^Z \) as $Z$ checks.
The specific code instances used in this paper appear in \Cref{tab:bbcodes}.
Each qubit is described by its type and a monomial label from
\begin{equation}
    M = \{x^p y^q \mid 0 \leq p < \ell, 0 \leq q < m\}.
    \label{eq:def_M}
\end{equation}
A $X$ check with label $\alpha \in M$ couples to $L$ data qubits with labels $A_i \alpha$ and to $R$ data qubits with labels $B_i \alpha$ for $i=1,2,3$; a $Z$ check labeled $\alpha$ couples to $L$ qubits labeled $B_i^{\top} \alpha$ and $R$ qubits labeled $A_i^{\top} \alpha$.
In total there are 12 categories of CNOT operations, denoted by triplets $(X/Z, L/R, i)$ with $i=1,2,3$.

\textbf{Syndrome extraction circuit and CNOT order.}
We retain the CNOT order in \Cref{eq:ibm_cnot_order}~\cite{bravyi_high-threshold_2024}.
\begin{equation}
    \begin{aligned}
    \text{Check qubits:} & \quad \text{X checks} & \text{Z checks} \\
     & & \text{Init} \ket{0} \\
    \text{Pass 1:} & \quad \text{Init} \ket{+} & \quad \text{CNOT}(R,Z,1) \\
    \text{Pass 2:} & \quad \text{CNOT}(X,L,2), & \text{CNOT}(R,Z,3) \\
    \text{Pass 3:} & \quad \text{CNOT}(X,R,2), & \text{CNOT}(L,Z,1) \\
    \text{Pass 4:} & \quad \text{CNOT}(X,R,1), & \text{CNOT}(L,Z,2) \\
    \text{Pass 5:} & \quad \text{CNOT}(X,R,3), & \text{CNOT}(L,Z,3) \\
    \text{Pass 6:} & \quad \text{CNOT}(X,L,1), & \text{CNOT}(R,Z,2) \\
    \text{Pass 7:} & \quad \text{CNOT}(X,L,3) & \text{Measure}\, Z \\
    & \quad \text{Measure}\, X & \text{Init} \ket{0} \\
    \text{Pass 1:} & \quad \text{Init} \ket{+} & \text{CNOT}(R,Z,1) \\
    ... & \quad ... & ... \\
    \end{aligned}
    \label{eq:ibm_cnot_order}
\end{equation}
The CNOT schedule in \Cref{eq:ibm_cnot_order} interleaves $X$- and $Z$-check operations to minimize depth while preserving the correctness of the syndrome extraction.
We refer to each row as a \emph{pass}; for example, ``CNOT($X$,$L$,2)'' in Pass 2 executes all CNOTs from every $X$ check to its corresponding $L$ data qubits specified by $A_2$ in parallel.
Passes 1 and 7 contain $n/2$ CNOTs each, while Passes 2 through 6 contain $n$ CNOTs.
A key objective of this work is to realize \textbf{maximum parallelism}, meaning \emph{all $n$ CNOTs within a pass execute simultaneously}.

\begin{table}[t]
    \centering
    \resizebox{0.7\columnwidth}{!}{%
    \begin{tabular}{c|c|c|c}
        \hline
        $[[n,k,d]]$ & \(\ell, m\) & \(A\) & \(B\) \\
        \hline
        $[[18,4,4]]$ & 3, 3 & \(x + 1 + y^2\) & \(y + 1 + x^2\) \\
        $[[72,12,6]]$ & 6, 6 & \(x^3 + y + y^2\) & \(y^3 + x + x^2\) \\
        $[[90,8,10]]$ & 15, 3 & \(x^9 + y + y^2\) & \(1 + x^2 + x^7\) \\
        $[[108,8,10]]$ & 9, 6 & \(x^3 + y + y^2\) & \(y^3 + x + x^2\) \\
        $[[144,12,12]]$ & 12, 6 & \(x^3 + y + y^2\) & \(y^3 + x + x^2\) \\
        $[[288,12,18]]$ & 12, 12 & \(x^3 + y^2 + y^7\) & \(y^3 + x + x^2\) \\
        \hline
    \end{tabular}
    }
    \caption{BB codes definition~\cite{wang_demonstration_2025,bravyi_high-threshold_2024}.}
    \label{tab:bbcodes}
\end{table}

\subsection{Theoretical bounds on qLDPC codes}

Theoretical results indicate that long-range interactions are not merely an implementation nuisance but a requirement for high-performance qLDPC codes. For local-expansion qLDPC codes implemented with 2D-local gates, the depth of a full syndrome-extraction cycle is at least \( \Omega(n/\sqrt{N}) \), where \(N\) is the total number of data, check, and ancilla qubits~\cite{delfosse_bounds_2021}. Thus one cannot simultaneously maintain constant encoding rate and constant-depth extraction using only \(O(n)\) qubits. Moreover, realizing a qLDPC code with distance \( d = \Theta(n^{1/2+\epsilon}) \) in 2D requires \( \Omega(n^{1/2+\epsilon}) \) interactions of range \( \tilde{\Omega}(n^{\epsilon}) \)~\cite{baspin_quantifying_2022}, implying that both the number and range of long-range interactions must grow with code size to surpass the \( d=\Theta(\sqrt{n}) \) regime. Taken together, these bounds motivate architectures and routing strategies that support many long-range operations efficiently, which is precisely the setting targeted in this work.

\subsection{Prior proposals for implementing BB codes}

\begin{table}[htbp]
    \centering
    \resizebox{\columnwidth}{!}{%
    \begin{tabular}{llll}
    \hline
    Design   & Platform & HW connectivity & Long-range interaction                \\ \hline
    Bravyi et al.~\cite{bravyi_high-threshold_2024} & SC       & Nearest neighbor+ & Long-range coupling      \\
    *Wang et al.~\cite{wang_demonstration_2025} & SC       & Nearest neighbor+ & Long-range coupling      \\
    Berthusen et al.~\cite{berthusen_toward_2024} & SC       & Stacked bilayer & Bell-pair generation \\
    Viszlai et al.~\cite{viszlai_matching_2024} & Atom     & All to all & AOD atom transport                  \\
    Poole et al.~\cite{poole_architecture_2024}  & Atom     & Finite radius & Rydberg blockade gate  \\
    This work     & SC       & Nearest neighbor & Oscillator network \\
    \hline
    \end{tabular}%
    }
    \caption{Proposals for implementing BB codes using superconducting circuits (SC) or neutral atom arrays (Atom).
    * denotes the experimental demonstration.}
    \label{tab:compare-layouts}
\end{table}

\begin{figure*}[htbp]
    \centering
    \includegraphics[width=\textwidth, trim={0 1em 0 0}, clip]{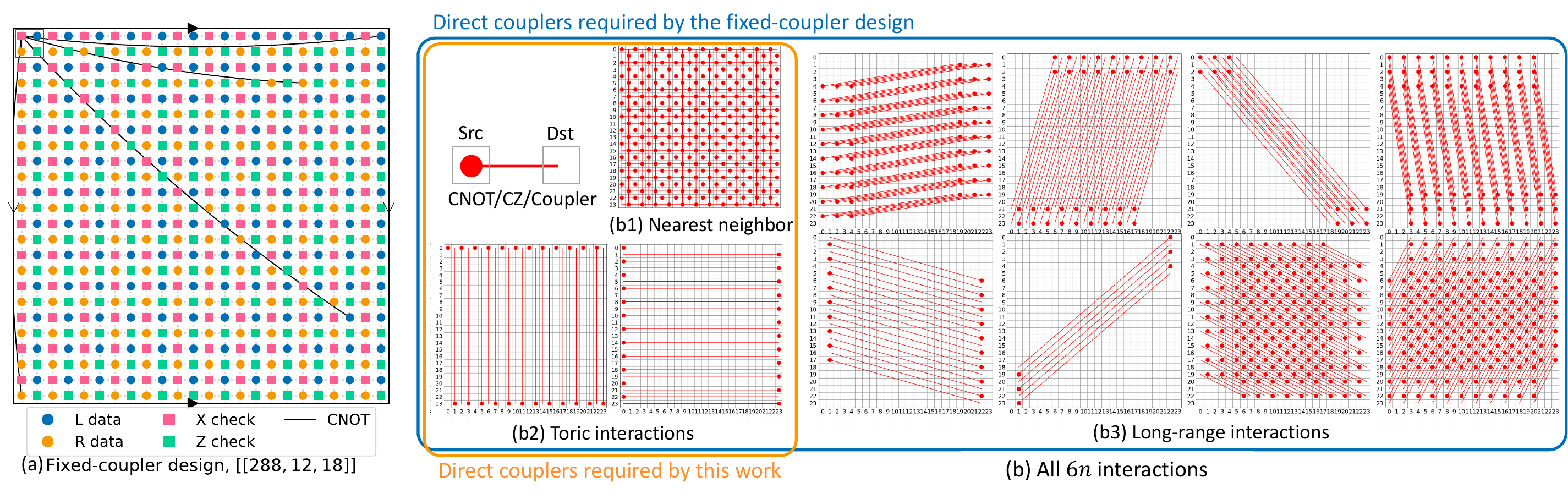}
    \caption{(a) Fixed-coupler layout. Each check qubit interacts with four nearest neighbors plus two long-range partners under periodic boundaries.
    Pairs of arrows with the same shape on the boundaries indicate the periodic boundary conditions.
    (b) There are \( 6n \) interactions for a BB code of size \( n \). 
    The blue box indicates the direct couplers that a fixed-coupler design must fabricate, whereas the orange box shows the much smaller set needed in our design; all other interactions are synthesized on demand by programming the communication network (\Cref{fig:programmable_comm_network}(a)). 
    }
    \label{fig:cmp_direct_and_programmable_couplers}
\end{figure*}

Implementation proposals for BB codes span multiple hardware platforms, summarized in \Cref{tab:compare-layouts}.
Neutral-atom systems offer flexible connectivity via acousto-optic deflector (AOD) transport~\cite{bluvstein_logical_2023, xu_constant-overhead_2023} and long-range interactions via Rydberg blockade~\cite{viszlai_matching_2024, walker_consequences_2007}, but face limitations in gate speed, error rates, and mid-circuit measurement~\cite{beverland_assessing_2022, lam_demonstration_2021, baker_exploiting_2021}.

For superconducting implementations, the tension between nearest-neighbor connectivity and non-local BB stabilizers has motivated three main strategies. The first is \textit{direct long-range coupling}: IBM's architecture~\cite{bravyi_high-threshold_2024} connects each qubit to four local neighbors plus two distant partners via long-range $c$-couplers under toric boundary conditions, which  requires $O(n)$ couplers whose lengths scale as $O(\sqrt{n})$. Mathews et al.~\cite{mathews_placing_2025} develop a multilayer place-and-route tool for this static-coupler setting. The second is \textit{long-range teleportation}: bilayer architectures~\cite{baumer_efficient_2023, berthusen_toward_2024} separate data/check qubits from entanglement generation, but require high-fidelity interlayer gates and rapid Bell-pair generation and distillation. The third is \textit{sequential SWAP routing}: common in NISQ compilation~\cite{li_tackling_2019}, but unattractive for large qLDPC codes because each transmon SWAP decomposes into three CNOTs and also propagates errors between the swapped qubits.

In parallel, software efforts explore richer instruction sets~\cite{zhou_louvre_2025}, trade long-range connectivity for deeper circuits~\cite{zhao_simple_2025}, or deform the code to limit interaction ranges~\cite{yang_planar_2025}. These limitations motivate architectures that preserve the standard BB-code schedule while supporting many long-range operations efficiently.

\subsection{Dual-rail qubit}

A dual-rail qubit~\cite{wu_erasure_2022, teoh_dual-rail_2023} encodes quantum information in the single-photon subspace of two superconducting microwave cavities. Through erasure detection~\cite{chou_superconducting_2024, graaf_mid-circuit_2024} the dominant photon-loss error becomes a heralded erasure, which can substantially improve QEC performance~\cite{wu_erasure_2022, mehta_bias-preserving_2025}. In practice, however, exploiting that benefit requires erasure-aware decoders; while such decoders exist for surface codes~\cite{wu_fusion_2023}, they are not yet available off the shelf for qLDPC codes~\cite{connolly_fast_2024, gokduman_erasure_2024, pecorari_quantum_2025, yao_cluster_2024}.

In the architecture studied here, dual-rail qubits interface naturally with the oscillator network (\Cref{sec:approach}) through beamsplitter SWAPs, and entangling operations are implemented with the Swap-Wait-Swap (SWS) gate~\cite{mehta_bias-preserving_2025}. Because this hardware stack has experimental performance benchmarks, it provides the basis for the realistic noise model and circuit-level simulations developed later in the paper. By chaining cavities, we can route photons across a 2D torus and execute long-range CZ gates between distant qubits, as detailed in \Cref{sec:error_modeling}.

\section{Programmable communication network}
\label{sec:approach}

\begin{figure*}[htbp]
    \centering
    \includegraphics[width=0.94\linewidth, trim={0 2em 0 0}, clip]{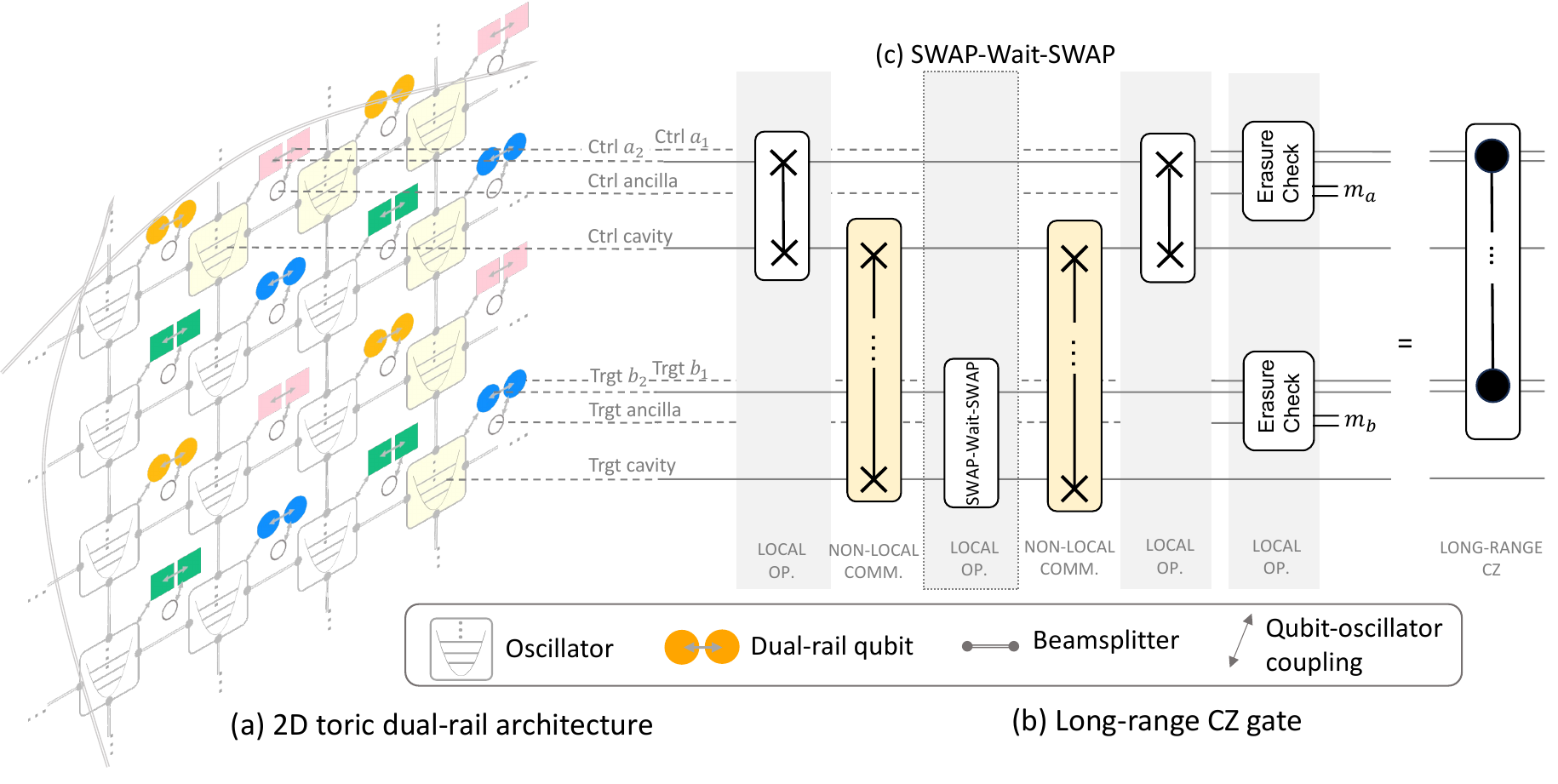}
    \caption{
    (a) Dual-rail+SWS implementation of a long-range CZ on a 2D toric oscillator network. 
    (b) Compilation into beamsplitter SWAPs and SWS gates.
    }
    \label{fig:os_network}
\end{figure*}

\textbf{Complexity of all-direct-coupler implementation.}
Implementing every long-range 2Q gate in the syndrome extraction via dedicated couplers is daunting.
\Cref{fig:cmp_direct_and_programmable_couplers}(b) represents each 2Q interaction as a pin and groups the $6n$ interactions for the $[[288,12,18]]$ code by parallelism so that pins within a group are parallel while different groups intersect.
This grouping mirrors network-on-chip intuition~\cite{kahng_vlsi_2011}: routing parallel couplers on the same metal layer is far easier than accommodating numerous crossings.
The fixed-coupler design~\cite{bravyi_future_2022, gambetta_surface_2025} relies on long-range couplers, but as both the interaction weight and range grow, maintaining high-fidelity couplers becomes increasingly difficult and risks degrading qubit coherence.

Scalability further exacerbates the challenge: the fixed-coupler design requires roughly \(2n\) couplers on a $\sqrt{n} \times \sqrt{n}$ lattice, so individual coupler lengths scale as \(O(\sqrt{n})\) as they span from one boundary to another.
Experimentally, only 32 long-range couplers have been demonstrated for the $[[18,4,4]]$ code, far short of what larger $n=144$ or $288$ BB codes would demand.

To overcome these bottlenecks we introduce a \textbf{programmable communication network}.
\Cref{fig:programmable_comm_network}(a) depicts the generalized architecture: a 2D torus of medium (e.g., cavities) forms a reconfigurable communication fabric, while dual-rail or transmon qubits at each site store the data and check states.
Fast tunable couplings between neighboring communication mediums dynamically transport quantum information across the torus, eliminating the need for fixed qubit-to-qubit long-range couplers.

A key benefit is the drastic reduction in fixed long-range couplers.
Only toric couplers need to be fabricated.
\Cref{fig:cmp_direct_and_programmable_couplers}(b) contrasts the two approaches: the blue box indicates the direct couplers that a fixed-coupler design must fabricate, whereas the orange box shows the much smaller set needed in our design; all other interactions are synthesized on demand by programming the communication network.

Multiple hardware realizations fit this blueprint.
\Cref{fig:programmable_comm_network}(b) illustrates an all-transmon variant in which qubits and the communication fabric are both transmons. 
Long-range SWAPs can be synthesized from three-CNOT transmon SWAPs, though this approach has low fidelity.
Since we are free to initialize the state of the medium to \( \ket{0} \), one CNOT can be omitted.
Alternatively, iSWAP and CZ gates can be used, requiring only two 2Q gates as well~\cite{krizan_quantum_2024}.
We take the latter synthesis as the representative example
of two-2Q gate transmon SWAPs and compare to other implementations.
\Cref{fig:programmable_comm_network}(c) depicts a hybrid oscillator-transmon scheme (also called Continuous-Discrete Variable, CV+DV)~\cite{liu_hybrid_2024} where oscillators carry information between transmon qubits.
This approach requires two full round-trips per long-range gate, resulting in high latency and decoherence; it has yet to be experimentally demonstrated.
\Cref{fig:programmable_comm_network}(d) shows the dual-rail + Swap-Wait-Swap (SWS) variant we emphasize in this work.
Here dual-rail qubits interface with the oscillator (cavity) via beamsplitter SWAPs, and entangling operations use high-fidelity SWS gates, yielding superior performance compared to the CV+DV option.
Because dual-rail + SWS has existing experimental benchmarks we can construct realistic noise models and run circuit-level simulations; these appear in \Cref{sec:error_modeling} and \Cref{sec:evaluation}.

\textbf{Separating data storage and communication.}
A key design goal of this architecture is to prevent long-range 2Q gates from propagating errors across the system. As shown in \Cref{fig:programmable_comm_network}(a), when quantum information travels through the communication mediums, we can decouple the qubit from the medium, effectively isolating data qubits from errors occurring in the communication network during long-range gate execution.

\textbf{Non-local gate implementation.}
Executing BB-code syndrome extraction efficiently demands compiling logical operations into hardware-native primitives and carefully scheduling the resulting non-local interactions.
In our architecture a long-range CZ decomposes into two ingredients (see \Cref{fig:os_network}(b)): beamsplitter SWAPs that shuttle the control-mode photon through the oscillator network and Swap-Wait-Swap gates that enact the CZ when the photon reaches the target.
The control photon swaps into the network, traverses to the target, participates in the SWS interaction, and then returns to the originating dual-rail qubit for erasure detection.

The CV+DV architecture follows a similar pattern but must perform two complete round-trips per long-range CNOT (four one-way traversals)~\cite[Figure~20(a)]{liu_hybrid_2024}, doubling the latency and increasing decoherence relative to the dual-rail approach.

The forthcoming \Cref{sec:layout_and_routing} introduces a qubit layout and routing strategy on the 2D torus.
Although motivated by dual-rail + SWS, the layout techniques generalize to any SWAP-based long-range implementation, including the all-transmon and CV+DV variants discussed above.

\section{Layout and routing co-design}
\label{sec:layout_and_routing}

In this section we detail the co-design of layout and routing for BB codes to achieve \textbf{maximum parallelism}, meaning all \( n \) CNOT gates required in a pass execute simultaneously.
Fixed-coupler layout keeps four of the six CNOTs local and routes the remaining two via long-range links under periodic boundary conditions, minimizing the number of non-local interactions.
Our programmable communication network instead favors six medium-range interactions per check.
\Cref{sec:evaluation} shows that this choice lowers coupler complexity and syndrome extraction cost.

Two symmetries are essential.
(1) \textbf{Translation invariance:} as illustrated in \Cref{fig:translation_invariant}(c), every interaction between a check and its data partners point in the same direction on the torus.
(2) \textbf{Cell invariance:} \Cref{fig:search_layout}(a,b) partitions all four qubit types into repeating $2\times 2$ cells whose structure tiles the layout.

Our strategy is therefore (a) to find a layout that preserves both symmetries while minimizing the maximum 2Q gate distance, and (b) to exploit those symmetries to route 2Q gates with maximal concurrency.

\subsection{Layout searching}

\begin{figure}[htp]
    \centering
    \includegraphics[width=\linewidth]{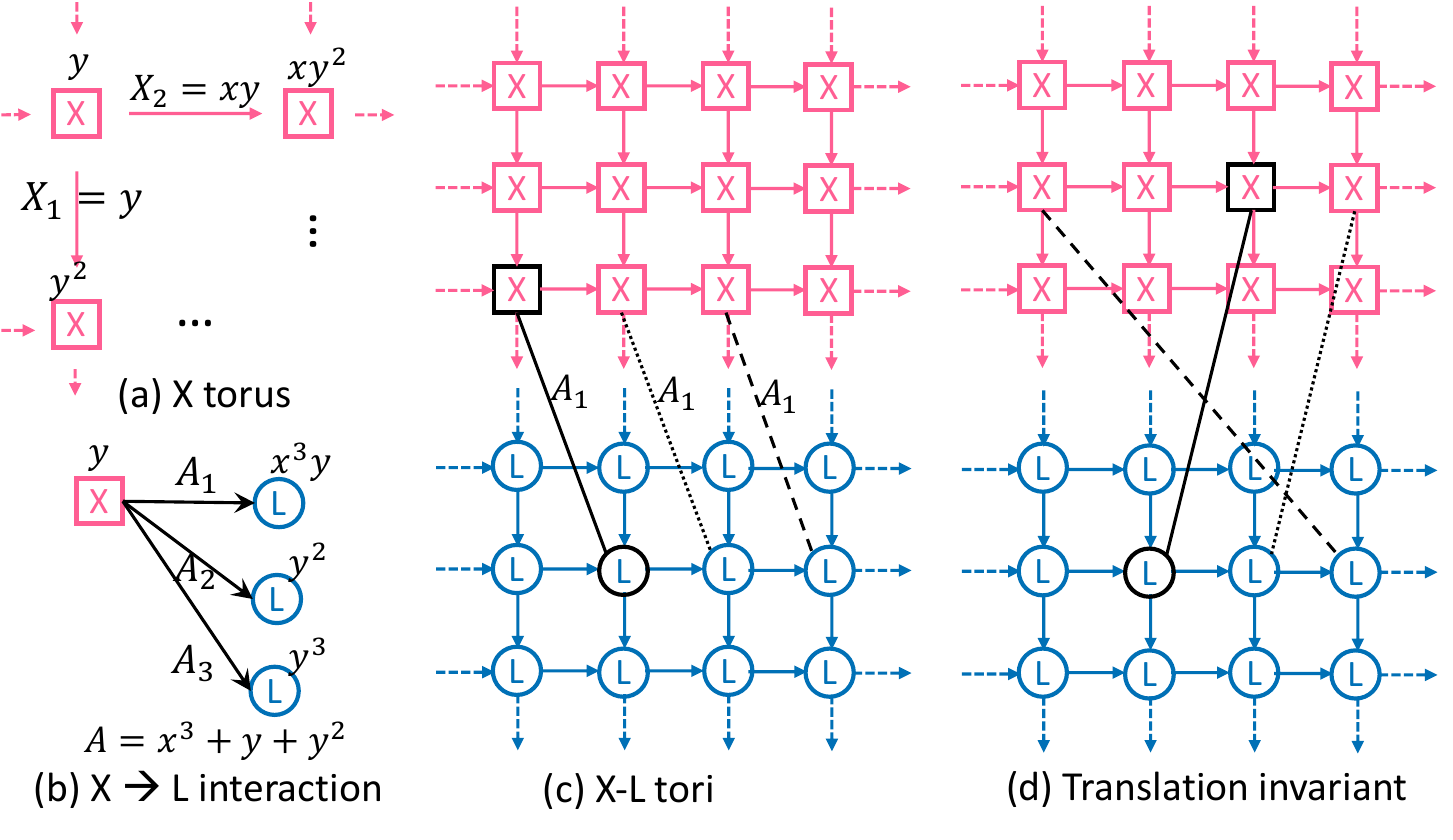}
    \caption{Translation invariance.
    (a) \( X_1,X_2 \) forms a torus for all the $X$ check qubits.
    (b) Interactions between $X$ and $L$ qubits.
    (c,d) After changing the relative offset of the two tori,
    all the interactions are still of the same direction on a $X$ torus.}
    \label{fig:translation_invariant}
\end{figure}

In the monomial representation each qubit is labeled by its type and an element of $M$ (defined in \Cref{eq:def_M}).
Each type admits a two-monomial basis generating all qubits of that type, e.g., \(\langle L_1,L_2\rangle = \langle X_1,X_2\rangle = \langle R_1,R_2\rangle = \langle Z_1,Z_2\rangle = M\), so every type forms a torus as in \Cref{fig:translation_invariant}(a).
Checks interact with data qubits according to the BB code definition; for instance, an $X$ check labeled \(\alpha\) couples to $L$ data qubits \(A_1\alpha, A_2\alpha, A_3\alpha\) as shown in \Cref{fig:translation_invariant}(b).

If we align the torus bases so that \(X_i=Z_i=L_i=R_i\) for \(i=1,2\), the layout becomes translation invariant.
For a representative $X$-$L$ interaction labeled by \(A_1\) in \Cref{fig:translation_invariant}(c), shifting the $X$ torus right by 2 and up by 1 leaves all interaction directions unchanged.
Formally, if the solid edge corresponds to \(A_1\alpha=\beta\), then the dotted edge corresponds to \(A_1(\alpha X_2)=\beta X_2=\beta L_2\), showing that torus translations preserve interaction directions.

\begin{figure}[htp]
    \centering
    \includegraphics[width=\linewidth]{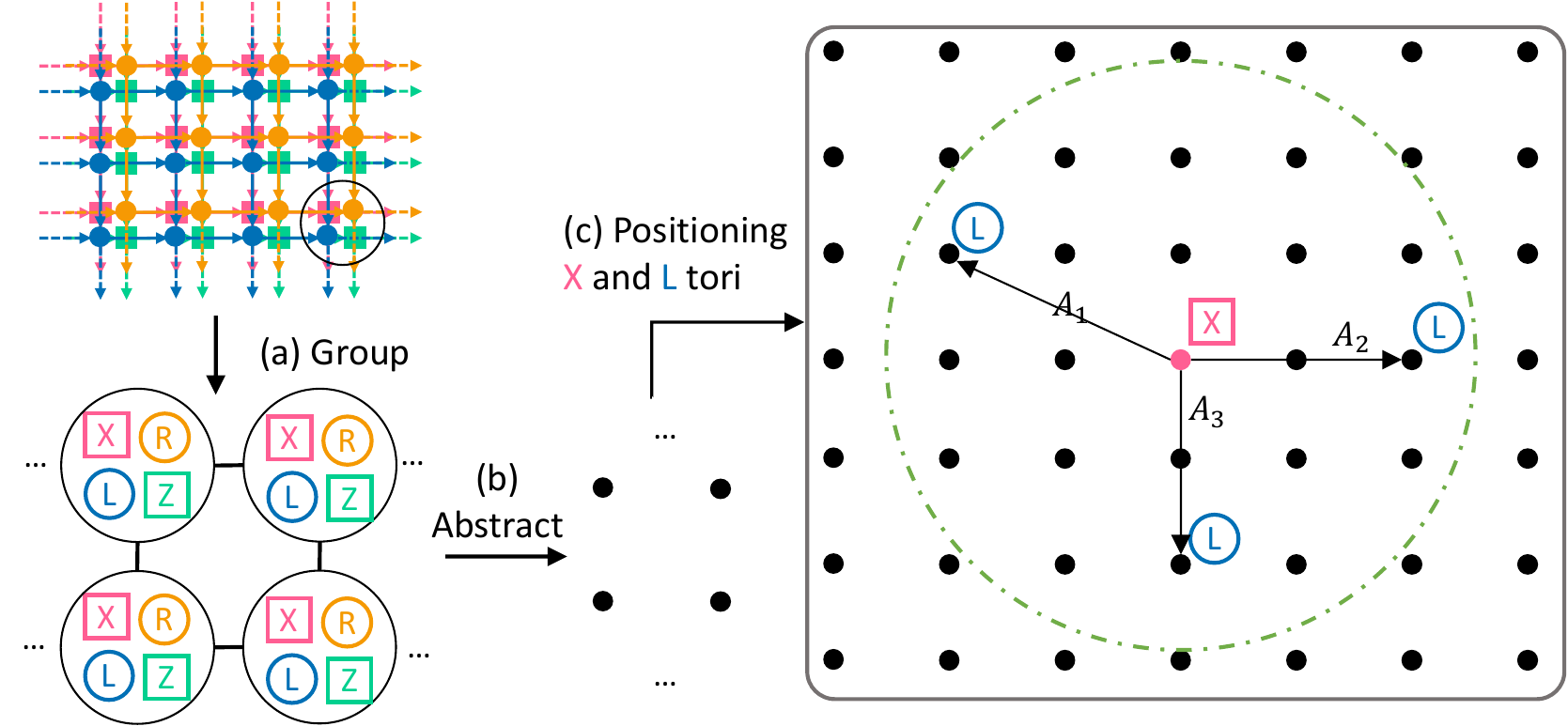}
    \caption{Search layout.
    (a,b) Stack the four types of tori on top of each other,
    and group them into 2 by 2 cells.
    (c) Make an abstraction of the cells by a dot.
    (d) Positioning $X$ and $L$ tori s.t. the distance of the interactions are less than \( D \).}
    \label{fig:search_layout}
\end{figure}

To minimize the maximum gate distance we follow the procedure in \Cref{fig:search_layout}: stack the four tori, group them into translation-invariant $2\times 2$ cells, and abstract each cell as a dot.
We then determine relative offsets between torus pairs.
For example, fix the $X$ torus so the central qubit carries label \(1\), implying its partner $L$ qubits have labels \(A_1, A_2, A_3\).
For every basis generating \(M\) we ask whether there exists a placement of the $L$ torus such that all three interactions lie within distance \(D\).
If none exists we increase \(D\) and retry.
After settling $X$-$L$ pair we repeat for $X$-$R$, $Z$-$L$, and $Z$-$R$ pairs.
Preserving translation invariance means a single check-qubit search suffices to determine each pairwise placement,
ensuring runtime efficiency.
For the \(n=756\) code this search completes within seconds on a single core of a MacBook Pro with an M1 Pro CPU.

\subsection{XY Routing on torus}

\begin{figure*}[htp]
    \centering
    \includegraphics[width=\linewidth, trim={0 1em 0 0}, clip]{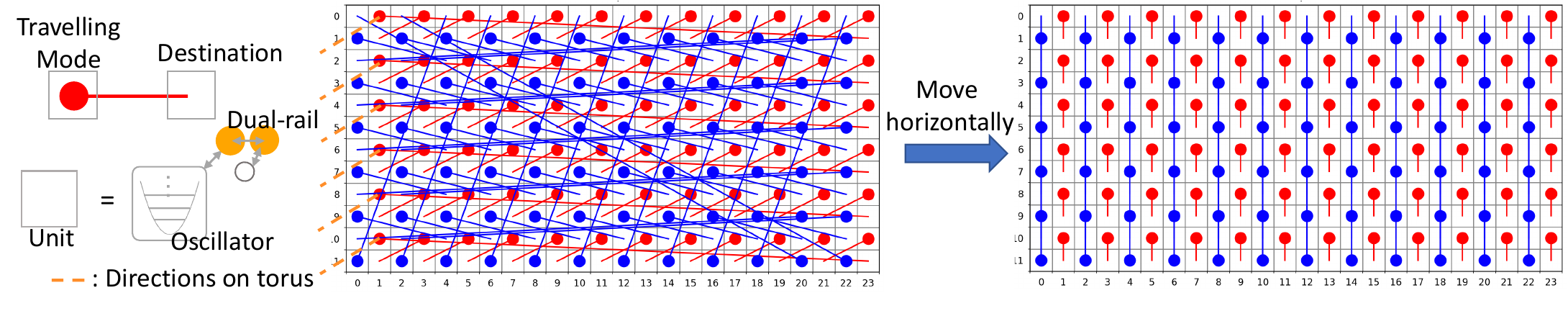}
    \caption{XY routing on torus.
    We first route the mode along the X axis (horizontal direction).
    The red pins (X checks) indicate the mode traveling to the left,
    while the blue pins (Z checks) indicate the mode traveling to the right.
    After X routing, all the pins become vertical,
    and we route on the Y axis.
    There are no conflicts during the routing process,
    because we preserve the translation invariance
    and cell invariance.
    }
    \label{fig:routing}
\end{figure*}

We route long-range interactions using toric XY routing~\cite{jerger_-chip_2022}.
Here the \textit{mode} (the photon residing in a cavity) first travels along the X direction, then along Y.
\Cref{fig:routing} depicts each 2Q interaction as a pin whose head shows the traveling mode and tail shows its destination on a grid where each box corresponds to an cavity with an attached dual-rail qubit.

Red pins (X-check routes) move leftward on the torus, blue pins (Z-check routes) move rightward, and dashed orange lines indicate their true wrapped paths on the torus.
Although the figure appears congested, translation invariance ensures all modes of the same color move coherently in one direction, so no conflicts arise.
During X routing, cell invariance guarantees there is at least one idle cavity between adjacent modes, preventing collisions.
After completing the X leg, all pins align vertically, allowing conflict-free Y routing.

\Cref{fig:routing} shows that every X and Z check (all \(n\) 2Q gates) in a given pass routes in parallel without conflict.
This saturates available resources: inserting any additional traveling mode would obstruct one of its four nearest neighbors.



\section{Gate noise modeling}
\label{sec:error_modeling}

\begin{figure}[htp]
    \centering
    \includegraphics[width=0.96\linewidth]{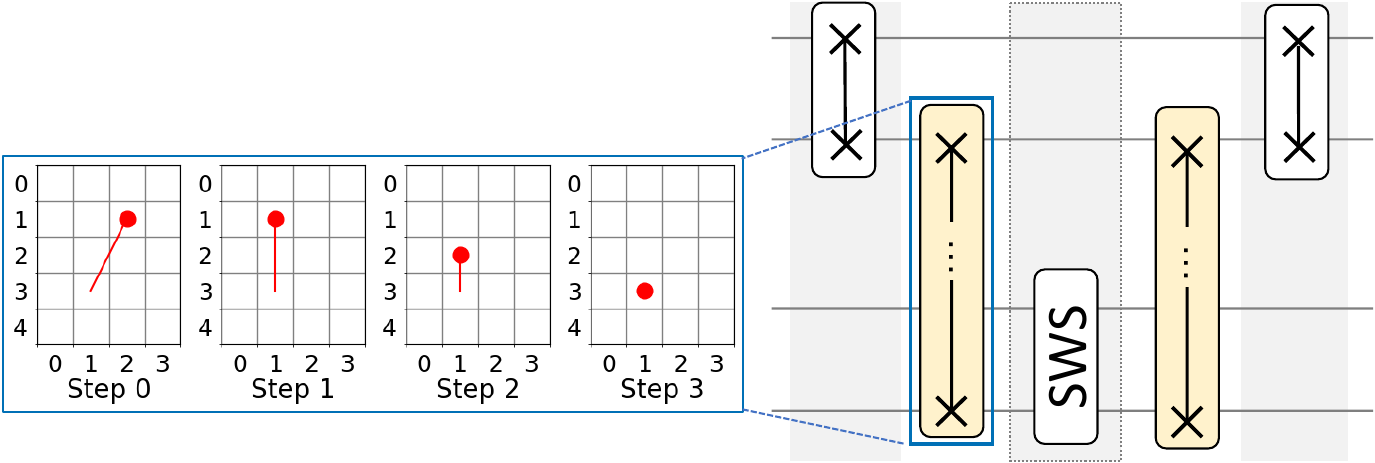}
    \caption{Define \( r \) in unit of the number of SWAP gates in one-way trip.
    This might include idling time, i.e., the mode is waiting in the cavity.}
    \label{fig:define_r}
\end{figure}

In this section, we model the error rate of long-range CZ gates using Swap-Wait-Swap (SWS) gates based on experimental data from previous component-level studies~\cite{mehta_bias-preserving_2025}, as summarized in \Cref{tab:hardware_params}
and \Cref{tab:dual_rail_qubit_properties}.

There are mainly following error mechanisms:
Photon loss during beamsplitter SWAP, idling and SWS gates;
residual dephasing and bit-flip errors;
and false positive rate of erasure detection.
We neglect no-jump backaction~\cite{teoh_dual-rail_2023, michael2016new, mehta_bias-preserving_2025, koottandavida_erasure_2024, teoh2023error}, a second-order error from unequal photon loss between the two logical basis states; assuming equal cavity lifetimes, the only asymmetry is that the excitation briefly populates the lossy coupler for one basis state but not the other, and this effect is subdominant to pure dephasing (see \Cref{app:nojump_backaction}).
We model the error rates as follows.

\textbf{Duration of long-range CZ gate}.
As shown in \Cref{fig:define_r},
we define \( r \) in unit of the number of SWAP gates in one-way trip.
This might include idling time, i.e., the mode is waiting in the cavity.
\begin{equation}
t_{\text{dur}}(r) = 2 \cdot t_{\text{SWAP}} \cdot (r+1) + t_{\text{SWS}} + t_{\text{det}}.
\end{equation}

\textbf{Photon loss in coupler}.
Only the mode from a control dual-rail qubit suffer from photon loss in the coupler.
\begin{equation}
    p_{\text{loss,cpl}}( t ) := 1 - e^{-t /  T_{1}^{\text{t}}}.
\end{equation}
We have designed in such a way that the control dual-rail is in \( \ket{+} \).
So, averaging over the two logical components gives an effective coupler duration
\( t = t_{\text{SWS}}/2 \), as derived in \Cref{app:effective_cavity_duration}.

\textbf{Photon loss in cavity}.
Both control and target dual-rail qubits suffer from photon loss in the cavity.
\begin{equation}
p_{\text{loss,cav}}(t) := 1 - e^{-t /  T_{1}^{\text{cav}}}.
\end{equation}
For target dual-rail qubits, $t=t_{\text{dur}}(r)$.
If the control dual-rail is in \( \ket{+} \),
averaging over the routed and non-routed logical components yields
\( t=t_{\text{dur}}(r) - t_{\text{SWS}}/2 \),
as derived in \Cref{app:effective_cavity_duration}. 

\textbf{Erasure rate.} Erasure rate is the probability that the erasure detection measurement
reports a photon loss, which is the sum of the true positive rate and the false positive rate
of erasure detection.
Any photon loss (even in the coupler) leads to a detectable error
and contributes to the true positive rate.
\begin{equation}
    \begin{aligned}
p_{\text{eras,ctrl}}(r) &=  p_{\text{loss,cav}}\left( t_{\text{dur}}(r)-t_{\text{SWS}}/2 \right) \\
&+ p_{\text{loss,cpl}}\left( t_{\text{SWS}}/2 \right) + p_{\text{FP}}
\end{aligned}
\end{equation}
\begin{equation}
p_{\text{eras,trgt}}(r) = p_{\text{loss,cav}}(t_{\text{dur}}(r)) + p_{\text{FP}}
\end{equation}
False negative rate
is proved to be negligible 
since the probability of missing an erasure is second-order
\cite{graaf_mid-circuit_2024}.

\textbf{Residual dephasing and bit-flip errors.}
Residual errors in the SWS entangling gate arise because erasure detection removes only excitation-loss events that leave the dual-rail codespace.
Both residual dephasing and bit-flip errors comprise two parts:
the intrinsic decoherence accumulated while traveling or idling in the cavity (first and second terms)
and the added error introduced by the SWS gate~\cite{mehta_bias-preserving_2025} (third term):
\begin{align}
    \label{eq:X_error_ctrl}
    p_{X,\text{ctrl}}(r) &= 1 - \exp\left( - \tfrac{t_{\text{dur}(r)}-t_{\text{SWS}}}{T_{\text{bit-flip,ctrl}}^{\text{DR}}} \right) + 2.8 \times 10^{-6} \\
    \label{eq:Z_error_ctrl}
    p_{Z,\text{ctrl}}(r) &= 1 - \exp\left( -\tfrac{t_{\text{dur}(r)}-t_{\text{SWS}}}{T_{\phi,\text{ctrl}}^{\text{DR}}} \right) + 3.9\times 10^{-4} \\
    \label{eq:X_error_trgt}
    p_{X,\text{trgt}}(r) &= 1 - \exp\left( - \tfrac{t_{\text{dur}(r)}-t_{\text{SWS}}}{T_{\text{bit-flip,trgt}}^{\text{DR}}} \right) + 0.5 \times 10^{-6} \\
    \label{eq:Z_error_trgt}
 p_{Z,\text{trgt}}(r) &= 1 - \exp\left( -\tfrac{t_{\text{dur}(r)} - t_{\text{SWS}}}{T_{\phi,\text{trgt}}^{\text{DR}}} \right) + 1.1 \times 10^{-4}
\end{align}

For comparison with other implementations (\Cref{sec:evaluation}),
we sum up these error rates
\begin{equation}
    \begin{aligned}
    p_{\text{2Q}}(r) &= p_{X,\text{ctrl}}(r) + p_{Z,\text{ctrl}}(r) + p_{X,\text{trgt}}(r) + p_{Z,\text{trgt}}(r) \\
    & + p_{\text{eras,ctrl}}(r) + p_{\text{eras,trgt}}(r).
    \end{aligned}
\end{equation}

\textbf{Circuit-level simulation.}
We first use layout and routing algorithm in \Cref{sec:layout_and_routing} to
derive the scheduling and duration of every long-range CZ gate.
Then, we calculate all the error rates above for each gate,
and use 
\verb|stim|~\cite{gidney_stim_2021} to perform \( d \) syndrome extraction cycles.
We use BP+OSD decoder~\cite{roffe_decoding_2020} to decode the error syndrome.

\section{Evaluation}
\label{sec:evaluation}

We evaluate the proposal bottom-up, examining fabrication cost, 2Q gate fidelity and latency, layout efficiency, routing performance, and overall QEC behavior.
For the noise-model validation we simulate the $[[18,4,4]]$ code, which has been experimentally realized.
Layout, routing, and scalability studies target the $[[144,12,12]]$ “gross” code and the $[[288,12,18]]$ code, both of which have logical-operation schemes projected to outperform surface code baselines.

\subsection{Fabrication cost}

\begin{figure}[htp]
    \centering
    \includegraphics[width=\linewidth]{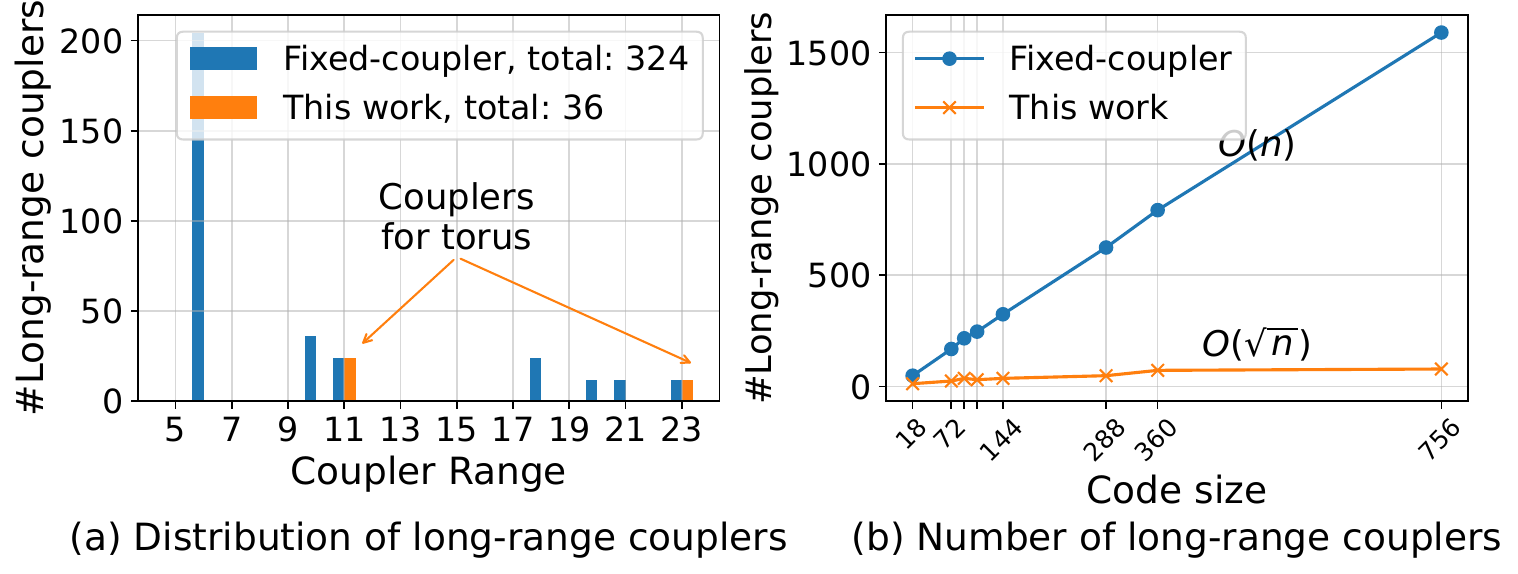}
    \caption{Comparison of the long-range couplers for the fixed-coupler design and our design. 
    }
    \label{fig:cmp_long_range_couplers}
\end{figure}

The fixed-coupler layout connects each check qubit to four nearest neighbors plus two long-range partners via fixed couplers under periodic boundaries.
Our architecture needs only toric couplers.
\Cref{fig:cmp_long_range_couplers}(a) shows the coupler-range distribution for the gross code: the fixed-coupler design needs 324 long-range couplers whereas we require just 36 (11\%).
As code size grows our coupler count scales as \(O(\sqrt{n})\) instead of \(O(n)\), as highlighted in \Cref{fig:cmp_long_range_couplers}(b).

\subsection{Layout performance}

\begin{figure}[htp]
    \centering
    \includegraphics[width=\linewidth, trim={0 1em 0 0}, clip]{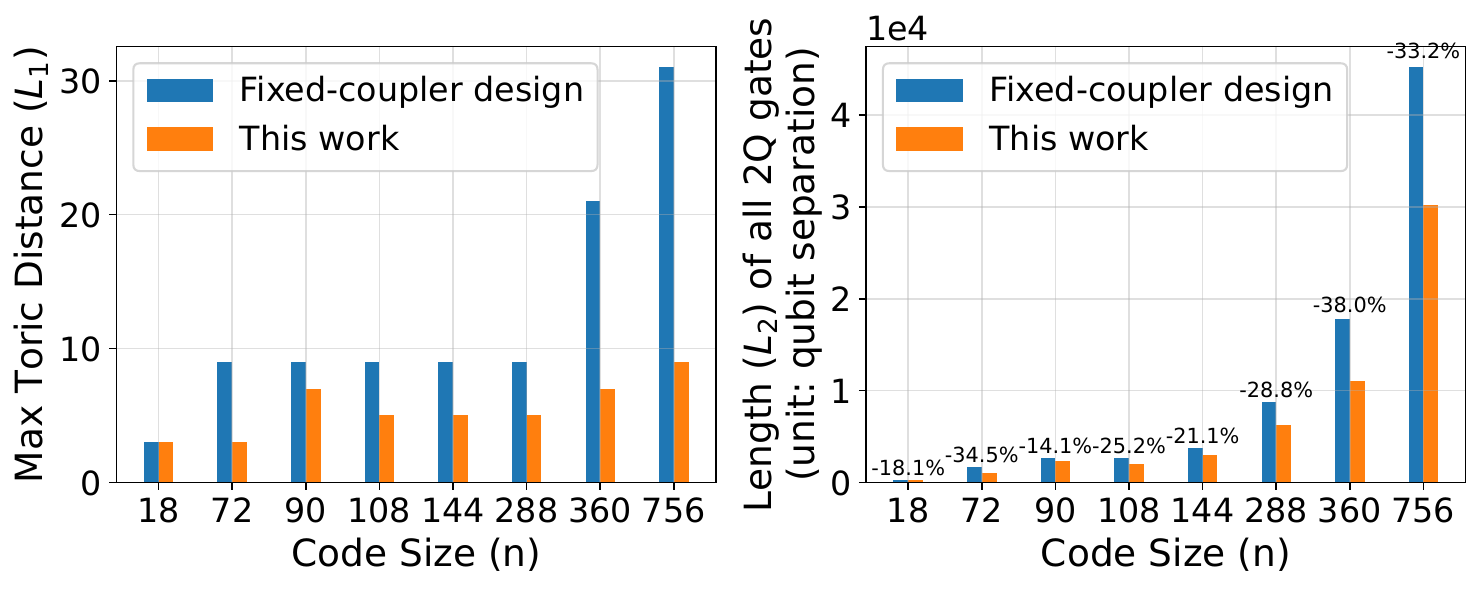}
    \caption{Comparison of the layout of the fixed-coupler design and our design.
    }
    \label{fig:cmp_length_on_layout}
\end{figure}
Compared to the fixed-coupler layout, our design markedly lowers the maximum toric distance (the \(L_1\) distance on the torus), especially for larger codes (\Cref{fig:cmp_length_on_layout}(a)).
Summing the \( L_2 \) lengths of every 2Q gate in a syndrome extraction cycle shows an 18--38\% reduction (\Cref{fig:cmp_length_on_layout}(b)), consistent with our preference for six medium-range gates over a mix of four local plus two very long ones.
Total wire length approximates the fabrication complexity of implementing couplers
or interactions.

\subsection{Layout and routing performance}

\begin{figure}[htp]
    \centering
    \includegraphics[width=\linewidth]{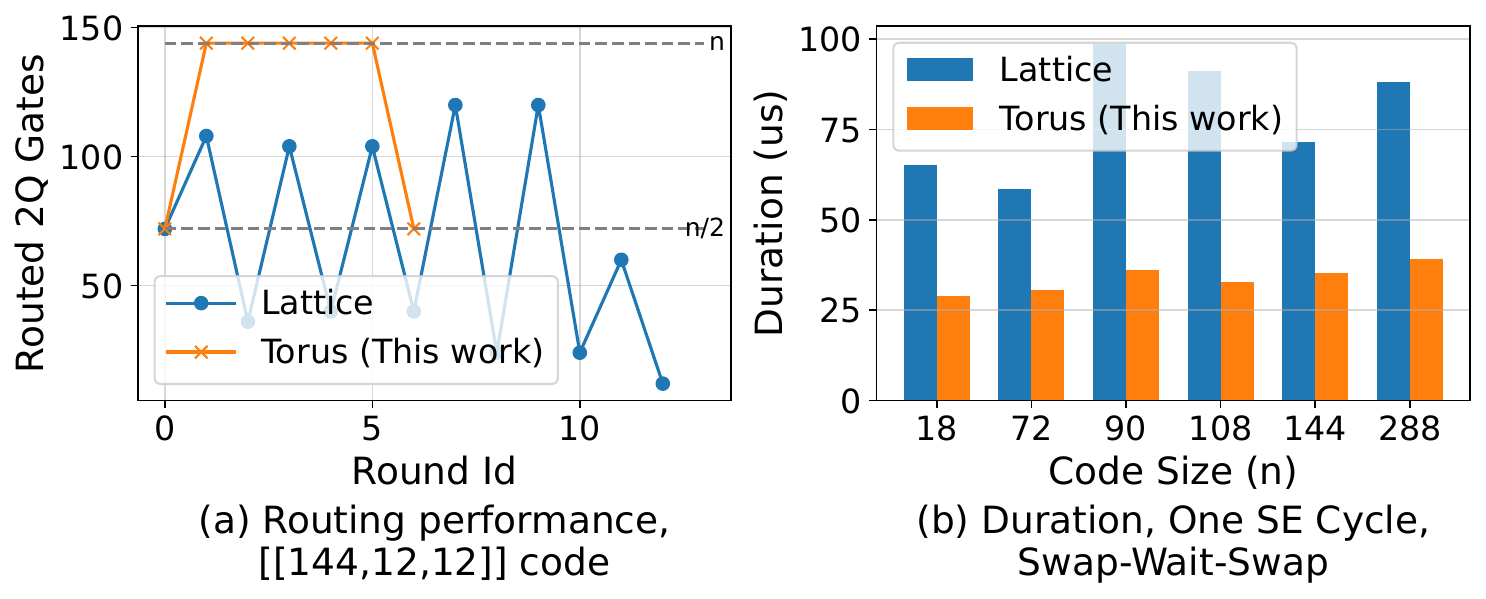}
    \caption{
    (a) The number of routed 2Q gates in each round of syndrome extraction.
    (b) The total duration of one syndrome extraction cycle, using dual-rail+SWS implementation.
    }
    \label{fig:lattice_vs_torus}
\end{figure}
We also benchmark routing with and without toric couplers.
For the lattice baseline we adopt the best-known layout with minimal maximum distance~\cite{poole_architecture_2024} and run a greedy XY routing algorithm.
Our design achieves maximum parallelism: orange line in \Cref{fig:lattice_vs_torus}(a) shows that each routing round matches the per-pass gate counts \(n/2, n, n, n, n, n, n/2\) from \Cref{eq:ibm_cnot_order}.
The lattice layout suffers many conflicts, roughly doubling the total duration (\Cref{fig:lattice_vs_torus}(b)), demonstrating that toric couplers plus code symmetries (\Cref{sec:layout_and_routing}) are key to maximum parallelism.

\subsection{2Q gate fidelity and duration}
\begin{figure}[htp]
    \centering
    \includegraphics[width=\linewidth, trim={0 1em 0 0}, clip]{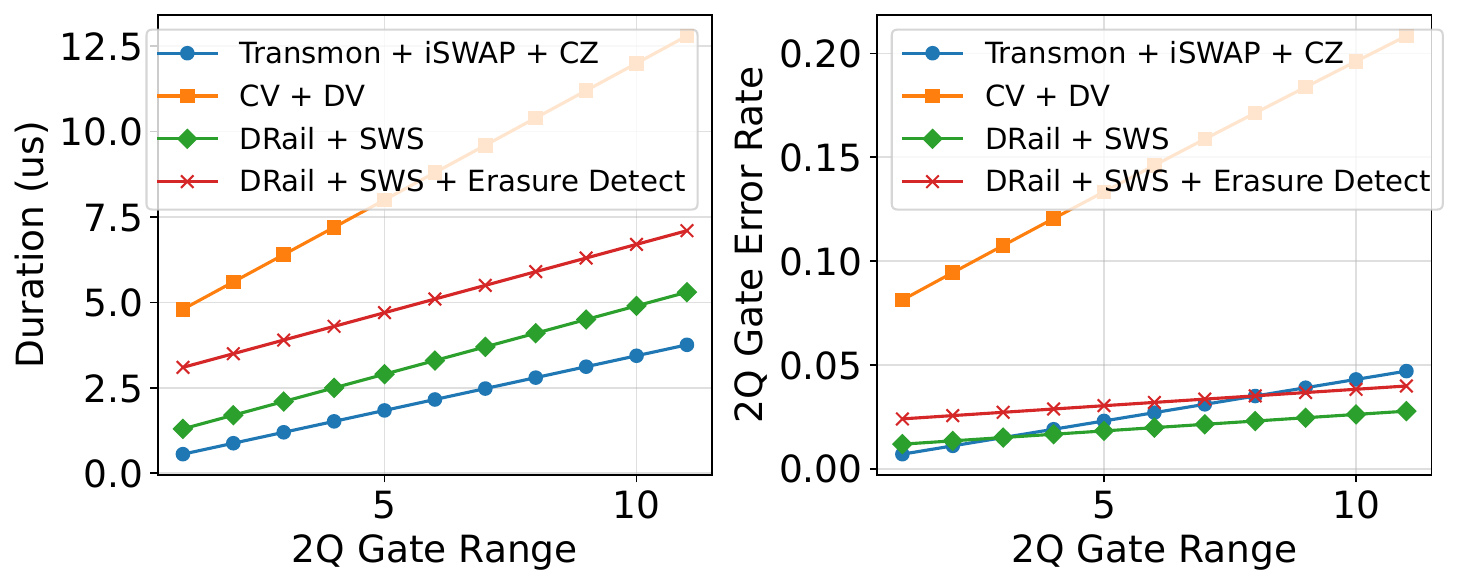}
    \caption{Comparison of the 2Q gate implementations.}
    \label{fig:cmp_2q_gate_impl}
\end{figure}
We compare several SWAP-based long-range 2Q gate implementations from \Cref{fig:programmable_comm_network}(b--d).
Because long-range gates might entail lengthy idles, we account for decoherence accumulated while the mode travels when calculating the error rate,
and prefer lower error rates over shorter durations.
Detailed error models for iSWAP+CZ and CV+DV implementations appear in \Cref{app:hardware_params}.
As shown in \Cref{fig:cmp_2q_gate_impl},
iSWAP+CZ is the fastest, whereas dual-rail+SWS exhibits the slowest error growth as range increases.
For dual-rail+SWS we consider variants with and without erasure detection, and later explain how this choice affects QEC performance.

\begin{figure}[htp]
    \centering
    \includegraphics[width=\linewidth]{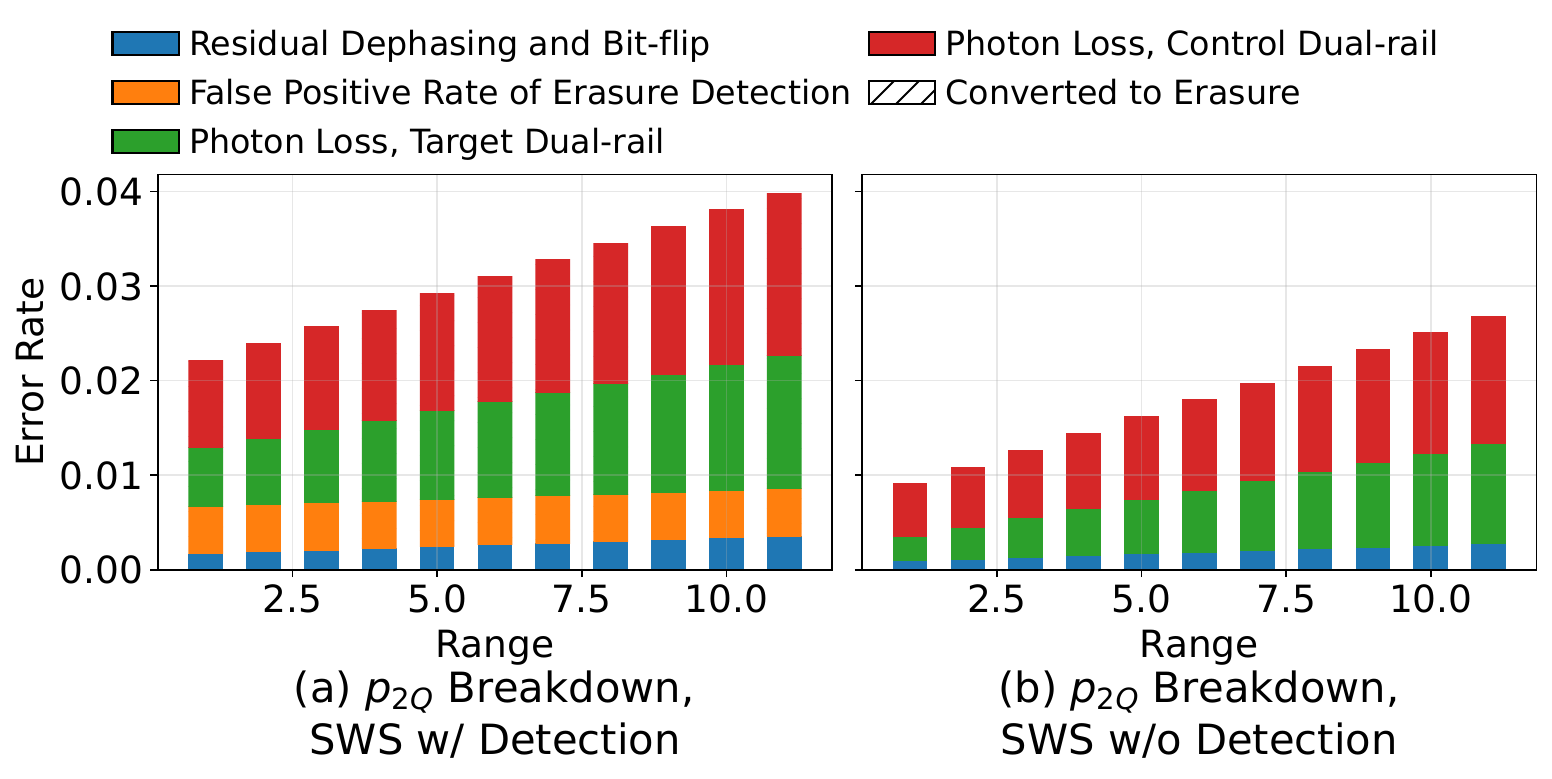}
    \caption{2Q gate error rate breakdown
    of dual-rail+SWS implementation, (a) with and (b) without erasure detection.
    Erasure detection converts photon losses into flagged erasures (hatched bars),
    while itself introduces latency and false positives, raising the total error rate.
    }
    \label{fig:p2q_breakdown}
\end{figure}
\Cref{fig:p2q_breakdown} decomposes the 2Q error rate.
While erasure detection converts photon losses into flagged erasures (hatched bars), the additional latency and false positives raise the total error probability, making erasure detection unattractive unless the decoder explicitly leverages erasure information. More explanation is in \Cref{subsec:qec_performance}.

\subsection{QEC performance}
\label{subsec:qec_performance}
\begin{figure}[htp]
    \centering
    \includegraphics[width=\linewidth, trim={0 1em 0 0}, clip]{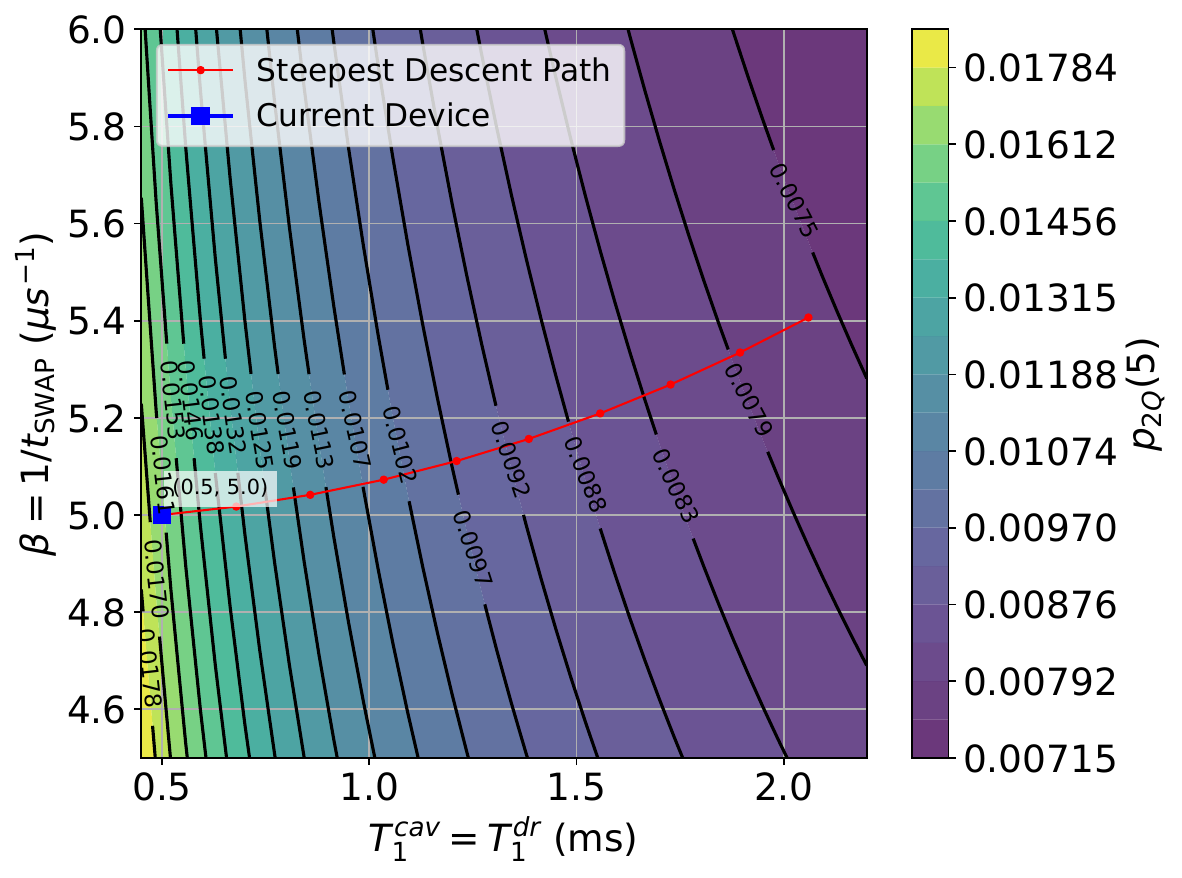}
    \caption{Contour plot of the worst-case \( p_{\text{2Q}} \) in the syndrome extraction circuit for the $[[18, 4, 4]]$ code as a function of \( t_{\text{SWAP}} \) and \( T_{1}^{\text{cav}} \).
    Start from current hardware parameters, we find a steepest descent path to mimic the future improvement,
    and sample \( (t_{\text{SWAP}}, T_{1}^{\text{cav}}) \) pairs along the path.
    We perform memory experiment simulation for each pair, shown in \Cref{fig:cmp_LER}.
    Since dual-rail qubits have only one photon,
    \( T_{1}^{\text{cav}} \) is roughly the same as \( T_{1}^{\text{dr}} \).
    }
    \label{fig:alpha_beta_landscape}
\end{figure}
\begin{figure}[htp]
    \centering
    \includegraphics[width=0.9\linewidth, trim={0 1em 0 0}, clip]{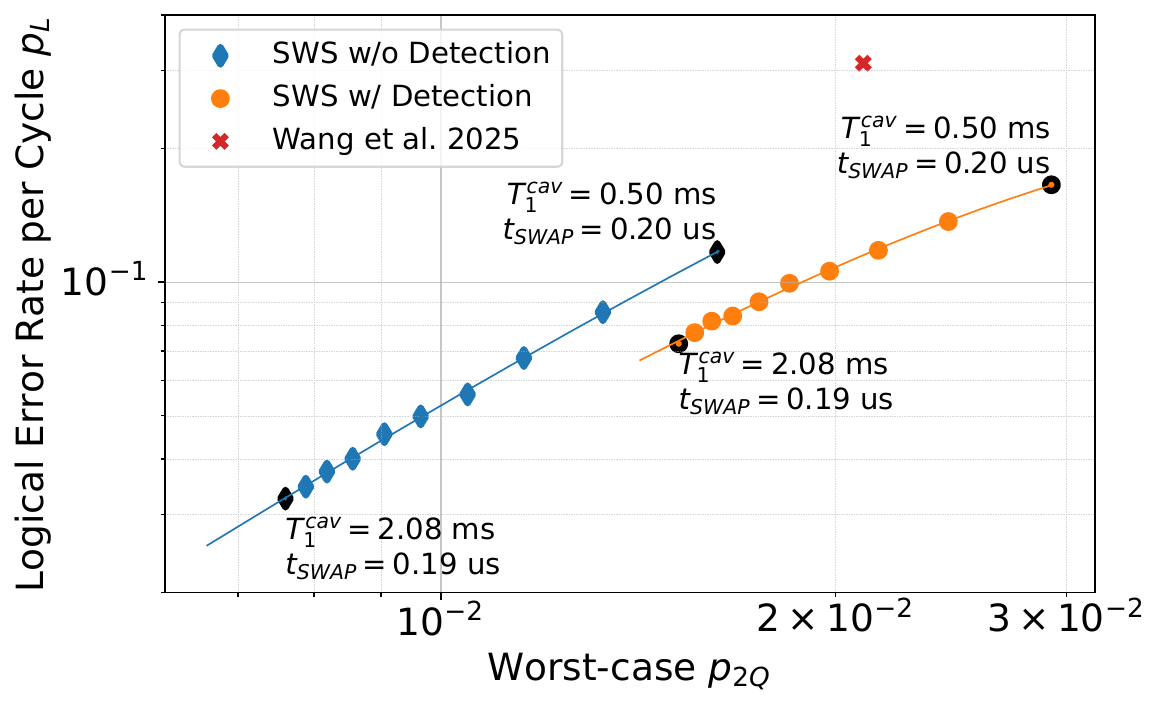}
    \caption{Comparison of the logical error rate of $[[18, 4, 4]]$ code.
    Hardware parameters are drawn from \Cref{fig:alpha_beta_landscape}.
    Red cross~\cite{wang_demonstration_2025} indicates \( p_L = 0.3115 \) and \( \epsilon_{k} = 0.0891 \),
    where \( p_L \) is the logical error rate per cycle of all the logical qubits,
    and \( \epsilon_{k} \) is the logical error rate per logical qubit per cycle.
    Rightmost point of the blue line indicates \( p_L = 0.1170 \) and \( \epsilon_{k} = 0.0306 \).
    }
    \label{fig:cmp_LER}
\end{figure}

\textbf{Define physical error rate.}
Surface code studies typically plot \(p_{\text{2Q}}\) versus logical error rate \(p_L\) because all 2Q gates are local and homogeneous.
BB codes mix gate lengths and error rates, so we instead plot the worst-case \(p_{\text{2Q}}\) (defined by the largest \(r\), including travel and idle time) against \(p_L\).
Varying \(p_{\text{2Q}}\) via future hardware improvements requires assumptions about which parameters can realistically advance; we focus on faster beamsplitter SWAPs \(t_{\text{SWAP}}\) and longer cavity lifetimes \(T_{1}^{\text{cav}}\).
Starting from current values we trace a steepest-descent path in the \((t_{\text{SWAP}}, T_{1}^{\text{cav}})\) plane (\Cref{fig:alpha_beta_landscape}), sample points along it, and run memory experiments for each to obtain \(p_L\) versus worst-case \(p_{\text{2Q}}\) for the $[[18,4,4]]$ code (\Cref{fig:cmp_LER}).

Following Wang et al.~\cite{wang_demonstration_2025} we also compute the logical error rate per logical qubit per cycle, \(\epsilon_k = 1 - (1 - p_L)^k\), to facilitate comparison with prior work.
The red cross in \Cref{fig:cmp_LER} marks their experimental result (\(p_L=0.3115\), \(\epsilon_k=0.0891\)).
Our simulated dual-rail+SWS system without erasure detection achieves \(p_L=0.1170\) and \(\epsilon_k=0.0306\) at current hardware parameters (rightmost point of the blue curve), a \(2.6\times\) improvement.

The erasure-detection curve in \Cref{fig:cmp_LER} lies to the right of the no-detection curve, indicating degraded performance.
Current erasure checks take 1.8~\(\mu\)s (roughly equivalent to nine beamsplitter SWAPs in duration) and have a false positive rate of 0.0051 (comparable to five SWAPs worth of error), so the overhead outweighs the benefit when using a BP+OSD decoder that lacks erasure awareness.
We expect erasure detection to become advantageous once paired with decoders tailored to flagged errors, as observed previously for surface codes~\cite{wu_erasure_2022, mehta_bias-preserving_2025}.

\section{Conclusion and future directions}
This paper introduced a hardware-software co-design for implementing quantum LDPC codes on 2D toric programmable communication networks.
By focusing on the long-range interactions required by BB codes and grounding the analysis in current hardware parameters, we showed via error modeling and circuit-level simulation that practical performance improvements are within reach.
We hope these results provide a roadmap for near-term demonstrations of high-rate qLDPC codes.
A central insight is that co-designing hardware, layout, and routing to exploit code symmetries yields maximum parallelism and markedly simplifies hardware implementation.

The following are some future work directions.
First, realizing the benefits of dual-rail erasure detection will require qLDPC decoders explicitly designed to ingest erasure information.
Second, our architecture might be useful for logical operations~\cite{breuckmann_fold-transversal_2024, moussa_transversal_2016,yoder_tour_2025, he_extractors_2025} demanding long-range interactions and dynamic connectivity.
Finally, reducing the detection cadence, e.g., checking after every few CZ gates, could save latency but demand new routing designs.

\begin{acknowledgments}
This work is supported primarily by the U.S. Department of Energy, Office of Science, National Quantum Information Science Research Center, Co-design Center for Quantum Advantage (C2QA) under Contract No. DE-SC0012704. YD acknowledges partial support by the National Science Foundation (under awards OSI-2435244, CCF-2312754 and CCF-2338063), by QuantumCT (under NSF Engines award ITE-2302908), by AFOSR MURI (FA9550-26-1-B036). External interest disclosure: RJS and YD are consultants and equity holders for D-Wave Quantum, Inc.
\end{acknowledgments}

\appendix



\section{Hardware parameters}
\label{app:hardware_params}

We list the hardware parameters used in this work.
We also explain the error modeling for iSWAP+CZ and CV+DV implementations.
\begin{table}[htbp]
    \centering
    \resizebox{\columnwidth}{!}{
    \begin{tabular}{lllll}
        \hline
        \textbf{Module} & \textbf{Description} & \textbf{Param} & \textbf{Value} \\
        \hline
        \multirow{2}{*}{Entangling coupler} & Relaxation time & $T_1^{\text{t}}$ & $70$ $\mu$s~\cite{mehta_bias-preserving_2025} \\
        & Pure dephasing time (Ramsey) & $T_\phi^{\text{t}}$ & $327$ $\mu$s~\cite{mehta_bias-preserving_2025} \\
        \hline
        Cavity & Relaxation time & $T_1^{\text{cav}}$ & $\sim500$ $\mu$s~\cite{mehta_bias-preserving_2025} \\
        \hline
        \multirow{1}{*}{Dual-rail} & Relaxation time & $T_1^{\text{dr}}$ & $\sim500$ $\mu$s~\cite{mehta_bias-preserving_2025} \\
        \hline
        \multirow{2}{*}{Erasure detection} & Detection time & $t_{\text{det}}$ & $1.8$ $\mu$s~\cite{graaf_mid-circuit_2024} \\
        & False positive rate & $p_{\text{FP}}$ & $0.51$\%~\cite{graaf_mid-circuit_2024} \\
        \hline
        \multirow{4}{*}{Gate durations} & Conditional displacement & $t_{\text{CD}}$ & $1.0$ $\mu$s~\cite{eickbusch_fast_2022} \\
        & ZZ interaction & $t_{\text{ZZ}}$ & $2.0$ $\mu$s~\cite{tsunoda_error-detectable_2023} \\
        & Beamsplitter SWAP gate & $t_{\text{SWAP}}$ & $0.2$ $\mu$s~\cite{mehta_bias-preserving_2025} \\
        & Swap-Wait-Swap gate & $t_{\text{SWS}}$ & $0.5$ $\mu$s~\cite{mehta_bias-preserving_2025} \\
        \hline
    \end{tabular}
    }
    \caption{Error model parameters used in noise modeling and circuit-level simulation.}
    \label{tab:hardware_params}
\end{table}

\begin{table}[htbp]
    \centering
    \resizebox{\columnwidth}{!}{
    \begin{tabular}{llll}
        \hline
        \textbf{Desc.} & \textbf{Param} & \textbf{Ctrl qubit} & \textbf{Trgt qubit} \\
        \hline
        Pure dephasing time (Echo) & \( T^{\text{DR}}_\phi \) & $4.0$ ms & $4.8$ ms \\
        \hline
        Bit-flip & \( T^{\text{DR}}_{\text{bit-flip}} \) & $520$ ms & $1100$ ms \\
        \hline
        End-of-the-line meas. error & \( p_{\text{meas}} \) & $2\times 10^{-4}$ & $2.3\times 10^{-4}$ \\
        \hline
        Individual gate error & \( p_{1Q} \) & $9.0\times 10^{-5}$ & $1.08\times 10^{-5}$ \\
        \hline
    \end{tabular}
    }
    \caption{Dual-rail qubit properties~\cite{mehta_bias-preserving_2025}.}
    \label{tab:dual_rail_qubit_properties}
\end{table}


\subsection{iSWAP + CZ long-range gate}
To simplify the analysis, 
we optimistically assume \( p_{\text{2Q}}(1) = 0.001 \) is the 2Q gate error rate of local iSWAP/CZ/CNOT gate
among transmon qubits and local gate duration \( t_{\text{2Q}}(1) = 80 \) ns,
though the experimental results~\cite{krizan_quantum_2024} are worse than these values.
Each SWAP takes one iSWAP and one CZ gate,
according to Figure 1(d) in~\cite{krizan_quantum_2024},
and we ignore single-qubit gate errors and durations.
The circuit is the same as \Cref{fig:os_network}(b),
while replacing SWS with CNOT.
\begin{equation}
    t_{\text{dur}}(r) = 2 \cdot 2 t_{\text{2Q}}(1) \cdot r + 2 t_{\text{2Q}}(1) + t_{\text{2Q}}(1)
\end{equation}
\begin{equation}
    p_{\text{2Q}}(r) = 2 \cdot 2 p_{\text{2Q}}(1) \cdot r + 2 p_{\text{2Q}}(1) + p_{\text{2Q}}(1)
\end{equation}
In this architecture,
the control qubit is SWAPed all the way to the target qubit,
so qubit coherence is incorporated into the error rate calculation.

\subsection{CV + DV long-range gate}
In this architecture,
the 2Q gate error rate comprises two parts:
(1) transmon decay and dephasing during idling and mode traveling;
(2) photon loss.

From~\cite[Figure~20(a)]{liu_hybrid_2024}, 
we can write the duration as
\begin{equation}
    t_{\text{dur}}(r) = 4 \cdot (t_{\text{CD}} + r \cdot t_{\text{SWAP}}).
\end{equation}

We model transmon decay and dephasing by 
applying Pauli twirling to the amplitude damping and dephasing channels, yielding 
a Pauli channel with probabilities~\cite{geher_reset_2024}:
\begin{equation}
\begin{aligned}
    & p_X(t)=p_Y(t)=\frac{1}{4}\left(1-e^{-t / T_{1}^{\text{t}}}\right) \\
    & p_Z(t)=\frac{1}{2}\left(1-e^{-t / T_{2}^{\text{t}}}\right)-\frac{1}{4}\left(1-e^{-t / T_{1}^{\text{t}}}\right)
\end{aligned}    
\end{equation}

For photon loss, we have
\begin{equation}
    p_{\text{loss}}(t_{\text{dur}}(r)) = 1 - e^{-t_{\text{dur}}(r) \cdot \bar{n} /  T_{1}^{\text{cav}}}
\end{equation}
where \( \bar{n} \) is the average photon number estimated to be \( 1 \).
The total 2Q gate error rate is the idling error rate from both control and target qubits, plus the photon loss error rate,
\begin{equation}
    p_{\text{2Q}}(t) = p_X(t) + p_Y(t) + p_Z(t) + p_{\text{loss}}(t),
    \quad t = t_{\text{idle}}(r)
\end{equation}

\subsection{Effective cavity duration}
\label{app:effective_cavity_duration}

This section explains why the cavity-loss model for the control dual-rail qubit
uses an effective cavity duration
\( t_{\text{dur}}(r) - t_{\text{SWS}}/2 \).
Following Mehta et al.~\cite{mehta_bias-preserving_2025},
the SWS gate acts on
$
    (a_1,a_2,c,b_1,b_2),
$
where \( a_1 \) and \( a_2 \) are the control dual-rail cavities,
\( c \) is the coupler,
and \( b_1 \) and \( b_2 \) are the target dual-rail cavities.
In our long-range realization, the routed path becomes
\[
    (a_1,\; a_2,\; d_1,\; d_2,\; \cdots,\; d_r,\; c,\; b_1,\; b_2),
\]
where \( d_1,\; d_2,\; \cdots,\; d_r \) are the network cavities.

Using the logical convention
    $\ket{0_L} = \ket{01}$ and $\ket{1_L} = \ket{10}$,
write the control state as
$
    \ket{\psi_{\text{ctrl}}}
    =
    \alpha \ket{10}_{a_1,a_2}
    +
    \beta \ket{01}_{a_1,a_2}.
$
Assume that \( a_2 \) is the routed rail.
Then only the \( \ket{01}_{a_1,a_2} \) component places the excitation on the routed rail, so
$
    p_{\text{route}} = |\beta|^2
$
is the probability that the control excitation traverses
$
    a_2 \to \cdots \to c
$
and then returns along the reverse path.
For the logical \( \ket{+} \) input used in the main text,
\( p_{\text{route}} = 1/2 \).

There are therefore two relevant cases for the cavity exposure of the control excitation.
With probability \( 1 - p_{\text{route}} \),
the excitation starts in the non-routed rail \( a_1 \),
never enters the coupler,
and remains in cavity modes for the full duration \( t_{\text{dur}}(r) \).
With probability \( p_{\text{route}} \),
the excitation starts in the routed rail \( a_2 \),
travels through the network,
and enters the coupler only during the SWS interval.
Its cavity exposure is therefore
$t_{\text{dur}}(r) - t_{\text{SWS}}$.
Averaging over these two cases gives
\begin{equation}
    \begin{aligned}
        t_{\text{eff,cav}}^{\text{ctrl}}
        &=
        (1 - p_{\text{route}})\, t_{\text{dur}}(r)
        +
        p_{\text{route}} \left( t_{\text{dur}}(r) - t_{\text{SWS}} \right) \\
        &=
        t_{\text{dur}}(r) - p_{\text{route}} t_{\text{SWS}}.
    \end{aligned}
\end{equation}
For the \( \ket{+} \) input,
\( p_{\text{route}} = 1/2 \),
so
$
    t_{\text{eff,cav}}^{\text{ctrl}}
    =
    t_{\text{dur}}(r) - t_{\text{SWS}}/2.
$

\subsection{No-jump backaction in the SWAP network}
\label{app:nojump_backaction}

This section analyzes no-jump backaction for the control dual-rail qubit during a long-range SWS gate and justifies its exclusion from the main error model.

No-jump backaction arises when the two logical basis states experience different effective decay rates~\cite{michael2016new, teoh2023error, koottandavida_erasure_2024}. In our protocol, $|1_L\rangle$ has its excitation in cavity $a_1$ throughout, while $|0_L\rangle$ routes its excitation through $a_2 \rightarrow d_1 \rightarrow \cdots \rightarrow d_r \rightarrow c$ before interacting with the coupler and returning. The resulting decay-rate imbalance between logical basis states is
\begin{equation}
\Delta \kappa\, t_{\rm dur}
=
(\kappa_{c} - \kappa_{a_1})\, t_{\rm SWS}
+
(\bar{\kappa}_{a_2,d} - \kappa_{a_1})\, t_{\rm cav},
\end{equation}
where $\bar{\kappa}_{a_2,d} = \frac{1}{r+1}\!\left(\kappa_{a_2} + \sum_{i=1}^{r} \kappa_{d_i}\right)$ is the average decay rate along the routing path and $t_{\rm cav} = 2\,t_{\rm SWAP}(r+1)$ is the total cavity dwell time. For an initial $|{+_L}\rangle$ state, the combined pure dephasing and no-jump infidelity is~\cite{teoh_dual-rail_2023}
\begin{equation}
    \varepsilon_{\rm NJ} \approx 
    \frac{1}{2}\gamma_\phi t_{\rm dur}
    +
    \frac{1}{16}(\Delta \kappa\, t_{\rm dur})^2.
\end{equation}
Because the excitation of $|0_L\rangle$ averages over $r+1$ cavities drawn from the same distribution as $a_1$, the cavity imbalance $\bar{\kappa}_{a_2,d} - \kappa_{a_1}$ has zero mean, so averaging over the cavity ensemble gives
\begin{equation}
\mathbb{E}[\varepsilon_{\rm NJ}]
=
\frac{1}{16}
\left[
(\kappa_c - \kappa_{a_1})^2 t_{\rm SWS}^2
+
\mathrm{Var}(\bar{\kappa}_{a_2,d} - \kappa_{a_1})\, t_{\rm cav}^2
\right].
\end{equation}
The coupler term is fixed by device parameters and independent of routing distance. The cavity variance term asymptotes to $\sigma^2 t_{\rm cav}^2$ at large $r$, while $t_{\rm cav}$ grows with $r$, so the cavity contribution grows similarly to ordinary idle dephasing.Both contributions are second order to the pure dephasing term already included in the main error model (Eqs.~\ref{eq:X_error_ctrl}--\ref{eq:Z_error_trgt}).

No-jump backaction can in principle be suppressed by a midpoint logical $X$ echo, which ensures both logical basis states accumulate equal exposure to each lossy mode. This cancellation applies to repeated local SWS gates with an even number of opposite-polarity operations before erasure detection~\cite{mehta_bias-preserving_2025}, but does not directly extend to our protocol, which performs an erasure check after every long-range gate. Achieving cancellation here would require at least two
long-range gates between erasure checks. Partial mitigation is also possible by
modulating the stationary excitation between $a_1$ and $a_2$ during routing;
we leave both optimizations to future work.


\bibliography{ref-Kun,ref-Evan}

\end{document}